# Quantum Oscillations in kagome metals CsTi$_3$Bi$_5$ and RbTi$_3$Bi$_5$

Zackary Rehfuss[1], Christopher Broyles[1], David Graf[2], Yongkang Li[3], Hengxin Tan[3], Zhen Zhao[4,5], Jiali Liu[4,5], Yuhang Zhang[4,5], Xiaoli Dong[4,5], Haitao Yang[4,5], Hongjun Gao[4,5], Binghai Yan[3], Sheng Ran[1]

[1] *Department of Physics, Washington University in St. Louis, St. Louis, MO 63130, USA*
[2] *National High Magnetic Field Laboratory, Florida State University, Tallahassee, Florida 32310, USA*
[3] *Department of Condensed Matter Physics, Weizmann Institute of Science, Rehovot 7610001, Israel*
[4] *Beijing National Center for Condensed Matter Physics and Institute of Physics, Chinese Academy of Sciences, Beijing 100190, PR China*
[5] *School of Physical Sciences, University of Chinese Academy of Sciences, Beijing 100190, PR China*
(Dated: January 24, 2024)

We report quantum oscillation measurements on the kagome compounds ATi$_3$Bi$_5$ (A=Rb, Cs) in magnetic fields up to 41.5 T and temperatures down to 350 mK. In addition to the frequencies observed in previous studies, we have observed multiple unreported frequencies above 2000 T in CsTi$_3$Bi$_5$ using a tunnel diode oscillator technique. We compare these results against density functional theory calculations and find good agreement with the calculations in the number of peaks observed, frequency, and the dimensionality of the Fermi surface. For RbTi$_3$Bi$_5$ we have obtained a different quantum oscillation spectrum, although calculated quantum oscillation frequencies for the Rb compound are remarkably similar to the Cs compound, calling for further studies.

**Introduction** The Kagome lattice is a rich playground for investigating fundamental physics and uncovering new, exotic phases of matter. Not only does this lattice display features such as flat bands, Van Hove singularities, Dirac points, Dirac cones, highly anisotropic Fermi surfaces and Fermi surface nesting, but it also hosts a plethora of exotic phases, including frustrated magnetism, quantum spin liquids, chiral spin states, and various topological phases.

Previous studies on kagome lattice materials include the observation of Weyl fermions in the ferromagnet Co$_3$Sn$_2$S$_2$ [1, 2], magnetic skymions in the noncollinear antiferromagnet Fe$_3$Sn$_2$ [3–5], flat bands and Dirac points in antiferromagnetic FeSn [6] and paramagnetic CoSn [7, 8], chiral spin textures in the noncollinear antiferromagnet Mn$_3$X (X=Sn, Ge) [9, 10], and the observation of large anomalous hall effects in many of these kagome systems [11]. More recently, new families of kagome lattice materials have been discovered such as the AM$_6$X$_6$ which have a large chemical tunability and shows an array of magnetic phases [12, 13]. Another family, Nb$_3$X$_8$ (X = Cl, Br, I), have a trigonally distorted breathing mode kagome lattice which have prominent isolated flat bands and are proposed to be possible Mott insulators or obstructed atomic insulators [14].

Another family that has attracted significant interest is the AV$_3$Sb$_5$ (A=K, Cs, Rb) system (the "135" compounds). This family has been a hot topic in recent years due to the observation of multiple competing phases such as superconductivity, charge and pair density waves, nematic ordering and a large anomalous hall effect, in a single material [15–35]. A recent discovery has found that replacing vanadium with chromium leads to a new compound, CsCr$_3$Sb$_5$, that exhibits multiple phases and becomes superconducting under applied pressure [36, 37]. These complex symmetry breaking ordered states are found to compete or intertwine with each other, much like the situation of the high temperature superconductors [38].

The recently discovered ATi$_3$Bi$_5$ (A=Rb,Cs) compounds represents another family that is isostructural to the AV$_3$Sb$_5$ compounds, but differ vastly in their emergent phenomena. Magnetic and electrical measurements have not revealed any magnetic or charge order down to the superconducting transition near 4.8 K [39, 40]. Therefore this system becomes a perfect candidate to study the intrinsic electronic structure of the kagome lattice that is not intermingled with any type of magnetic or charge ordering. First principles calculations and photoemission measurements have identified essential features of the band structure due to the kagome lattice, including flat band, Dirac nodal loops and nodal lines [41–43]. Another way to measure the Fermi surface is to measure Quantum oscillation frequencies. Quantum oscillation frequencies have been calculated for ATi$_3$Bi$_5$, which demonstrate a strong 3D signature despite their cylinder-like Fermi surface geometry [44]. The Quantum oscillation measurements have previously been carried out, in low magnetic field, which only revealed relatively low frequencies [44–46]. In this work we report quantum oscillation measurements on both CsTi$_3$Bi$_5$ and RbTi$_3$Bi$_5$, in magnetic field up to 41.5 T. For the Cs compound, our results reveal 4 frequencies above 2000 T, consistent with DFT calculations. The Rb compound has different frequency spectrum from that of the Cs compound, which is likely due to a different Fermi energy.

**Methods** The samples were grown using a self flux method outlined elsewhere [39]. Due to the air sensitive nature of the ATi$_3$Bi$_5$ compounds, special handling was taken to minimize oxidization. To store the samples long term, they were stored in liquid nitrogen inside a Dewar. While transporting the samples, they were sealed in quarts ampules that were filled with argon gas at 0.33 Bar. Before the experiments, the samples were exposed to air briefly for mounting them onto the experiment platforms. The samples were optically inspected and samples were selected with nominal thickness under 100um and a surface area of roughly 200 $\mu$m x 200 $\mu$m, then they were encapsulated in GE varnish on the experimental platforms. Quantum oscillation measurements were performed at the high magnetic field laboratory in Tallahassee Florida using a tunnel diode oscillator (TDO) technique [47, 48] and a piezoresistive torque cantilever technique in fields up to 41.5 T and temperatures down to 350 mK. The torque and TDO signal were background subtracted using a combination of Gaussian filtering and a 3rd order polynomial fit. All the analysis of quantum oscillation data was carried out using Matlab 2023a with the curve fitter and signal analyzer toolboxes.

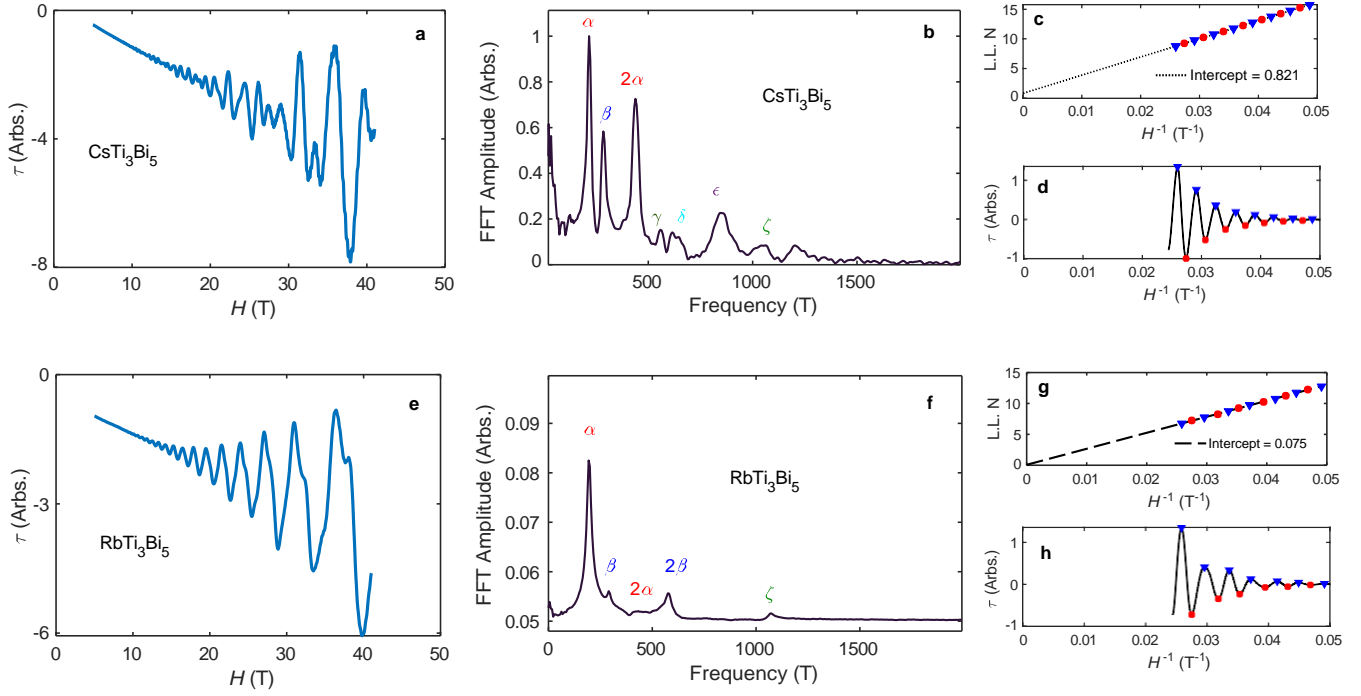

Figure 1. **(a,e)** Magnetic torque vs magnetic field (H) for $CsTi_3Bi_5$ and $RbTi_3Bi_5$ respectively. **(b,f)** Fourier transform amplitude vs frequency at $\theta = 0$ degrees for $CsTi_3Bi_5$ and $RbTi_3Bi_5$ respectively, with labels showing fundamental frequencies and harmonics. **(c,g)** Landau fan diagram showing Landau Level N vs inverse field for the $\alpha$ frequency peak for $CsTi_3Bi_5$ (c) and $RbTi_3Bi_5$ (g). Magnetic torque vs inverse field (1/H) with a band-pass filter set between 200 T and 250 T to isolate out the $\alpha$ frequency contribution to construct the Landau fan diagram for for $CsTi_3Bi_5$ (d) and $RbTi_3Bi_5$ (h).

For the density functional theory (DFT) calculations, crystal structures of $CsTi_3Bi_5$ and $RbTi_3Bi_5$ are fully relaxed with the Vienna *ab-initio* Simulation Package (VASP) [49], where the Perdew-Burke-Ernzerhof [50] type generalized gradient approximation is employed to mimic the electron-electron interaction. The cutoff energy for the plane-wave basis set is 300 eV, and a $k$-mesh of $12\times12\times6$ is used to sample the Brillouin zone. The DFT-D3 vdW correction [51] is considered in the structural relaxation. After that, the Full Potential Local Orbital (FPLO) software [52] is used to fit the band structure for the tight-binding Hamiltonians of both materials with the wannier basis set composed of Ti $d$ and Bi $p$ orbitals. The spin-orbital coupling and a $k$-mesh of $12\times12\times6$ are employed. Fermi surfaces of $CsTi_3Bi_5$ and $RbTi_3Bi_5$ are calculated from the tight-binding Hamiltonians. The detailed method for quantum oscillation frequency and phase calculation based on the Fermi surface can be found in Refs. [44, 53].

**Results** To study the Fermi surface in the $ATi_3Bi_5$ (A=Cs,Rb) family, we measured magnetic torque ($\tau$) and TDO frequency as a function of magnetic field strength, angle and temperature. Fig. 1(a) and Fig. 1(e) shows the torque signal measured as a function of magnetic field at 350 mK for $CsTi_3Bi_5$ and $RbTi_3Bi_5$ respectively. Clear quantum oscillations can be seen across the entire field range for both compounds, with the Cs compound showing more apparent frequency components than the Rb compound in the raw data. Fig. 1(b) and Fig. 1(f) show the discrete Fourier transform (FFT) of the background subtracted torque signal vs inverse field for the Cs and Rb compounds for the applied field perpendicular to the kagome plane ($\theta$=0).

We observe multiple frequency peaks for frequencies below 1500 T in both compounds. Some frequency peaks might be harmonics. To identify harmonic peaks, we check the ratio with fundamental frequencies across all observable angles and see if this ratio remained constant and an integer (Fig. S3). For the Cs one, there are multi-ple peaks with similar frequencies, making it difficult to distinguish between fundamental peaks and harmonics for some frequencies. With the assistance of DFT calculations (discussed later), we identified 6 fundamental frequency peaks at 216 T ($\alpha$), 284 T ($\beta$), 557 T ($\gamma$), 630 T ($\delta$), 850 T ($\epsilon$) and 1060 T ($\zeta$), and a suspected harmonic peak at 438 T ($2\alpha$). The frequency spectra for Rb torque measurements appear to be different from that for Cs one, even excluding harmonic peaks. In the Rb Sample we observed 3 fundamental frequency peaks at 193 T ($\alpha$), 290 T ($\beta$) and 1070 T ($\zeta$), and 2 suspected harmonic peaks at 430 T ($2\alpha$) and 580 T ($2\beta$). In the high frequency range, we observed a frequency peak around 4500 T for Rb shown in Fig. S4. This frequency peak was only clearly observed at 1 angle ($\theta$=32 degrees), and was very broad. Due to the broadness of the peak and it only being observed at one angle, this peak may not correspond to real quantum oscillations.

Fig. 1(c) and Fig. 1(g) show the Landau fan diagram made for the lowest frequency peak for the Cs and Rb compound respectively. The lowest frequency peak was isolated in the raw data using a band-pass filter and the signal is plotted vs inverse field (Fig. 1(d) and 1(h)). The maxima were assigned to n+3/4 and and minima were assigned to n+1/4. Then the intercept was calculated by extrapolating the linear maxima and minima to zero in inverse field. We obtain a intercept of 0.821 for the Cs compound and 0.075 for the Rb compound.

In order to gain insight into the effective mass of the charge carriers in this system, we measured the temperature dependence of quantum oscillation data. The FFT for the Cs compound is shown in Fig. 2(a) at various temperatures from 350 mK to 20 K. We see a clear monotonic decrease in the amplitude of the FFT indicating that all the frequency peaks originate from real quantum oscillations. The temperature dependence of the normalized FFT amplitude is shown in Fig. 2(b) for a few selected fundamental frequencies. The effective masses were calculated by fitting the temperature dependence of the FFT amplitudes using the standard Lifshitz-Kosevich formalism [54, 55] following eqn (1).

$$A(T) \propto \frac{\frac{am^*T}{\mu_0 H}}{sinh(\frac{am^*T}{\mu_0 H})}, \quad (1)$$

where $m^*$ is the effective mass, $T$ is the temperature, $H^{-1} = \frac{H_{min}^{-1} + H_{max}^{-1}}{2}$ ($\mu_0 H$=8.9 T) is the average inverse magnetic field for the analysis field range and $a$ is defined as 14.69 T/K. The calculated effective masses, in the unit of electron mass, are 0.115 ($\alpha$), 0.107 ($\beta$), 0.188 ($\epsilon$), 0.396 ($\zeta$). In this analysis, we did not include the frequencies that are close to harmonic peaks. Fig. 4(a) shows the temperature dependence of the FFT for the torque signal of Rb compound for $\theta$ = 40 degrees. The effective masses for the fundamental frequency peaks are 0.133 ($\alpha$), 0.098 ($\beta$), 0.130 ($\zeta$), The first two peaks $\alpha$ and $\beta$ being similar to the Cs compound, and the third peak $\zeta$ differing significantly.

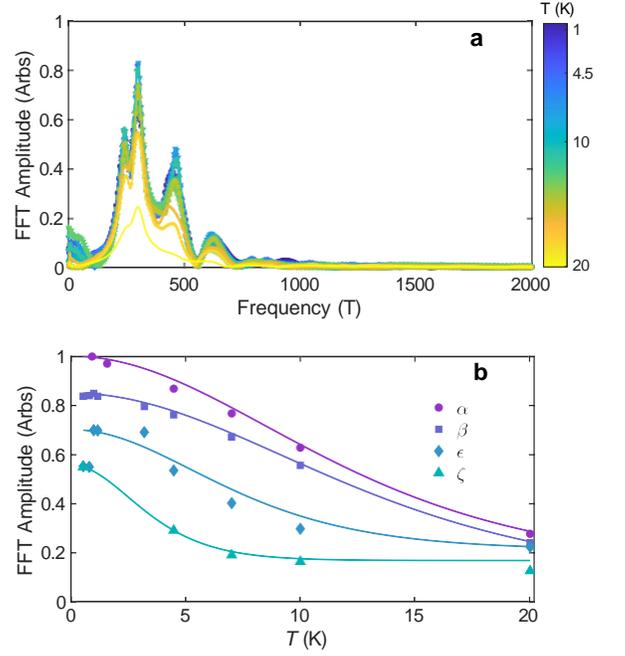

Figure 2. **(a)** Temperature dependence of the Fourier transform amplitude vs frequency, at $\theta$ = 40 degrees, of the background subtracted torque signal vs 1/H. for CsTi$_3$Bi$_5$. **(b)** Lifshitz-Kosevich fitting for the temperature dependence of the Fourier transform amplitudes for the observed fundamental frequency peaks.

To study the geometry of the Fermi surface, the angular dependence of the torque signal of Cs compound was measured and is shown in Fig. 3(a) for various angles between 0 and 70 degrees. All the peak frequencies monotonically increase with the angle between the magnetic field and the c-axis. At angles above 40 degrees, we see most of the peaks have begun to or already have vanished in amplitude, which is a good indicator of a 2D Fermi surface. We fit the peak frequency as a function of angle to 1/cos($\theta$), shown in Fig. 3(b). We see good agreement with the 1/cos($\theta$) fit at angles less than 30 degrees, but see a deviation from the 1/cos($\theta$) fit for higher angles, indicating the Fermi surface is not exactly two dimensional. Similar results were obtained for the Rb compound, as shown in Fig. 5.

Most of the frequencies observed in our torque data have been reported in the previous studies [44–46]. In order to get higher frequencies that were not previously observed, we also used a Tunnel Diode Oscillator (TDO) technique to study the quantum oscillations, which is more sensitive to the electric conductivity than magnetic susceptibility. We observed clear but small amplitude quantum oscillations in the raw TDO signal shown in Fig. S5. The data was analyzed in two ranges, a full field range from 5 to 41 T and a high field range from 20-30 T up to 41 T. The full field range was used to study the low frequency (less than 2000 T) oscillations, and the high field range was used to study the high frequency oscillations ($f > 2000$ T). One of the striking differences between the torque and the TDO techniques is the number of high frequency peaks observed in the TDO data. We were able to observe the same 6 fundamental peaks the torque revealed in the low frequency range, and an additional 4 fundamental frequency peaks above 2000 T, with clear temperature dependence, shown in Fig. S10 ($\theta = 40$ deg). The effective masses for all the 10 fundamental frequencies observed at $\theta = 40$ deg in TDO data are displayed in Table 1 with all seeming to be similar to those obtained from torque data. Angular dependence shows that these high frequency peaks also corresponds to quasi 2D Fermi surface pockets, shown in Fig. 6g.

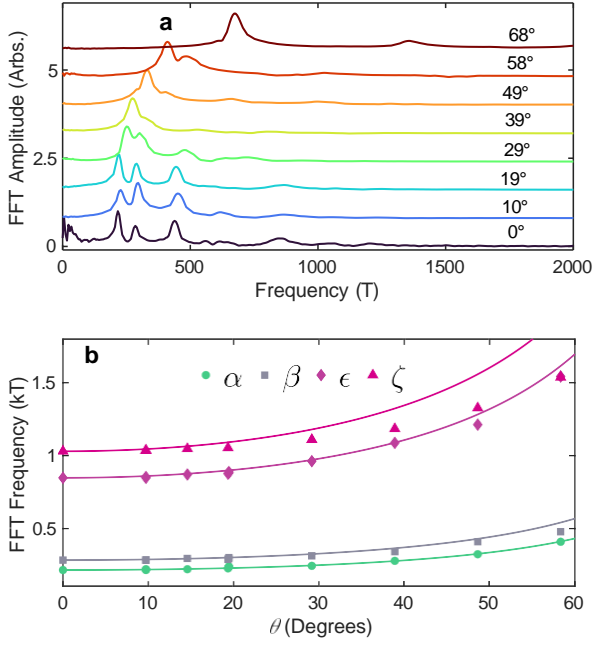

Figure 3. **(a)** Angular dependence of the Fourier transform amplitude (offset for clarity) vs frequency of the background subtracted torque signal vs 1/H. for the CsTi$_3$Bi$_5$ torque measurement. **(b)** 1/cos($\theta$) fitting for the angular dependence of the Fourier transform frequency in the for the observed fundamental frequency peaks.

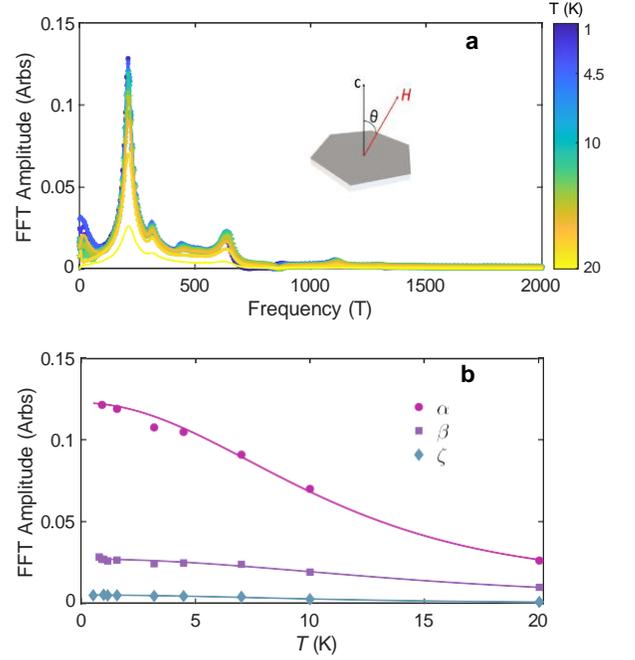

Figure 4. **(a)** Temperature dependence of the Fourier transform amplitude vs frequency, at $\theta = 40$ degrees, of the background subtracted torque signal vs 1/H. for RbTi$_3$Bi$_5$. The inset shows how we measure the angle relative to the ab plane. **(b)** Lifshitz-Kosevich fitting for the temperature dependence of the Fourier transform amplitudes for the observed fundamental frequency peaks in RbTi$_3$Bi$_5$.

**Discussions.** Having established the Fermi surface from experiments, we now explore its origin from the DFT calculations. Band structures of both materials including spin-orbital coupling (SOC) are shown in Fig. 7(a)&(b). Characteristic kagome band features are seen, including van Hove singularities at the $M/L$ points (e.g., at 0.2 eV), flat bands along the $M-K/L-H$ lines near $-0.5$ eV, and Dirac points at the $K/H$ points (e.g., at about 0.5 eV) that get gapped out by SOC. Notice that there are also plenty of Dirac points along other high symmetry $k$ paths, among which those on the $\Gamma$-M and $A-L$ lines are type II Dirac crossings. These Dirac crossings form Dirac nodal lines in the high-symmetry planes due to the symmetry constraint [53]. These calculations are in excellent agreement with recent angle-resolved photoemission spectroscopy (ARPES) measurements [41]. There are four bands that cross the Fermi level, giving rise to five pieces of Fermi Surface. Fig. 7(c)&(d) show the calculated Fermi surfaces for CsTi$_3$Bi$_5$ and RbTi$_3$Bi$_5$ respectively. Notice that only the green Fermi surface around the $M - L$ line is a hole pocket while all others are electron pockets.

| | $F_{theory}$(T) | $F_\tau$(T) | $F_{TDO}$(T) | $m_{theory}$ | $m_\tau$ | $m_{TDO}$ |
|---|---|---|---|---|---|---|
| $\alpha$ | 213 | 216 | 219 | -0.24 | 0.115 | 0.471 |
| $\beta$ | 336 | 284 | 277 | -0.22 | 0.107 | 0.522 |
| $\gamma$ | 542 | 557 | 438 | 0.22 | | 0.305 |
| $\delta$ | 713 | 643 | 593 | 0.26 | | 0.440 |
| $\epsilon$ | 802 | 848 | 838 | 0.24 | 0.188 | 0.545 |
| $\zeta$ | 889 | 1041 | 1019 | 0.32 | 0.396 | 0.474 |
| $\eta$ | 4569 | | 4601 | 0.72 | | 0.720 |
| $\theta$ | 4907 | | 4937 | 0.78 | | 0.735 |
| $\iota$ | 7488 | | 7575 | 1.62 | | 1.218 |
| $\kappa$ | 8111 | | 8360 | 1.68 | | 1.141 |

Table I. Calculated and measured values of the extremal orbits of the fermi surface of CsTi$_3$Bi$_5$ at $\theta = 0$ degrees. Column 1 is the calculated frequencies from DFT. Columns 2 and 3 are the measured frequencies from the torque and TDO measurements respectively. Column 4 is the calculated effective mass ratio $\frac{m}{m_e^*}$. Columns 5 and 6 are the measured effective masses for the torque and TDO measurements respectively.
4

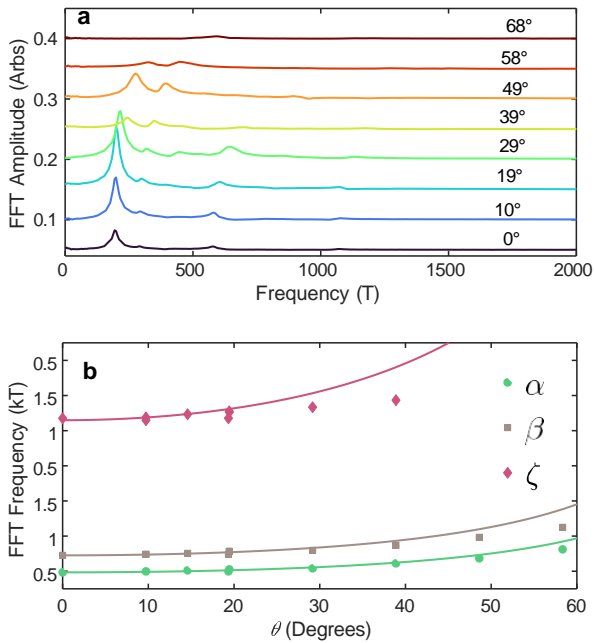

Figure 5. **(a)** Angular dependence of the Fourier transform amplitude (offset for clarity) vs frequency of the background subtracted torque signal vs 1/H. for RbTi$_3$Bi$_5$. **(b)** 1/cos($\theta$) fitting for the angular dependence of the Fourier transform frequency for the observed fundamental frequency peaks in RbTi$_3$Bi$_5$.

For the Cs compound, the calculation predicts 10 frequencies to be observed at $\theta$=0 degrees shown in column 1 of Table 1. (see details in [44]). We break these up into two ranges for our analysis; six "low frequencies" ($f < 2000$) and 4 "high frequencies" ($f > 2000$). Our TDO data revealed six low frequencies and four high frequencies (see Fig. 6(c) and column 3 Table 1), which sounds in good agreement with the calculations. In our torque data in Fig. 1(b), there are 7 observed low-frequency peaks, including fundamental peaks and harmonics. After excluding the harmonics following the procedure in the supplementary information, we are able to identify all the six low frequencies, shown in column 2 of Table 1. We point out that the calculated frequencies show some notable differences from our observations, which might be related to the Fermi energy difference. To further verify this scenario, we calculated the quantum oscillation frequencies as a function of the Fermi energy (Fig. S11). Fig. 8 shows the comparison between calculated frequencies at different Fermi energies and our measured ones. The measured frequencies fall within reasonable error bars from calculations at varied Fermi energies.

Because of the similarity in both crystal and electronic structures of RbTi$_3$Bi$_5$ and CsTi$_3$Bi$_5$ in theory, the calculated quantum oscillation frequencies for the Rb compound are remarkably similar to the Cs compound. However, we have observed a lot fewer frequencies in the Rb compound than in the Cs compound. Given the clear quantum oscillation signal of the Rb compound, this difference is unlikely related to the sample quality. One possibility is that the Fermi surface of the Rb compound might be very different from that of the calculation due to, such as doping, which can lead to different quantum oscillations. More studies are needed to elucidate the origin of the difference between Cs and Rb compounds.

Fig. 1(c) and 1(g) show the Landau fan diagrams for the lowest frequency peak for the Cs and Rb compounds respectively. The intercept of these plots is normally related to the topology of the Fermi surface. We have obtained an intercept of 0.821 for the Cs compound and 0.075 for the Rb compound, which seems to indicate a different topology of the Cs and Rb compounds and also aligns with the different oscillation spectra of these two materials. As suggested by recent theoretical studies [44, 53], the quantum phase shift in materials with both inversion and time-reversal symmetries can only be one or zero. Thus, we speculate that the Fermi pocket with the lowest frequency in the Cs compound (both inversion and time-reversal symmetric) has a non-trivial quantum phase shift (0.821$\pi$ is close to $\pi$) while that of the Rb compound has no phase shift (0.075$\pi$ can be regarded as zero). However, we could not rule of other possible reasons, such as sample variation. More importantly, as suggested by the studies [44, 53], we could not simply infer or exclude the existence of a non-trivial Berry phase in these Fermi pockets from the measured intercepts, because the quantum oscillation phase is composed of many different phases and the quantum phase shift is not a simple manifestation of the Berry phase.

To summarize, we report quantum oscillation measurements on the kagome lattice materials $ATi_3Bi_5$ (A=Cs,Rb) using tunnel diode oscillator and piezoelectric torque cantilever techniques for fields up to 41T. For Cs compound, we have observed all the 10 frequencies predicted by the DFT calculations, validating the Fermi surface of this family of kagome compound. The Rb compound has a different quantum oscillation spectrum, requiring further studies.

Research at Washington University was supported by the National Science Foundation (NSF) Division of Materials Research Award DMR-2236528. Research at the National High Magnetic Field Laboratory NHMFL was supported by NSF Cooperative Agreement No. DMR-2128556 and the State of Florida. H.Y. and H.G. are supported by grants from the National Natural Science Foundation of China (61888102), the National Key Research and Development Projects of China (2022YFA1204100), and the Chinese Academy of Sciences (XDB33030100). X.D. is supported in part by the the National Key Research and Development Pro-gram of China (2022YFA1403900), the Strategic Priority Research Program of Chinese Academy of Sciences (XDB33010200). B.Y. acknowledges the financial support from the European Research Council (ERC Consolidator Grant "NonlinearTopo", No. 815869) and ISF - Singapore-Israel Research Grant (No. 3520/20).

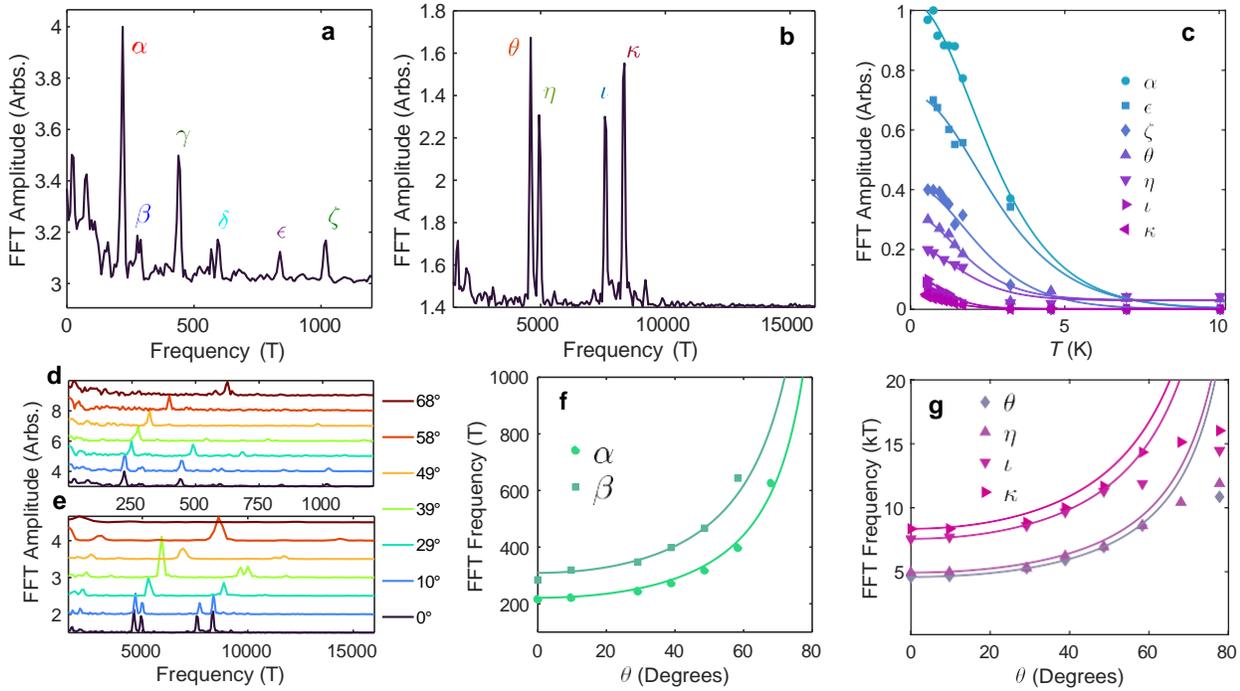

Figure 6. **(a,b)** The Fourier transform amplitude vs frequency, at $\theta = 0$ degrees, of the background subtracted TDO signal vs $1/H$. for $CsTi_3Bi_5$ in the low and high frequency range respectively with peak labels. **(c)** Lifshitz-Kosevich fitting for the temperature dependence of the Fourier transform amplitudes for the observed fundamental frequency peaks for the $CsTi_3Bi_5$ TDO measurement. **(d,e)** Angular dependence of the Fourier transform amplitude (offset for clarity) vs frequency of the background subtracted torque signal vs $1/H$. for the $CsTi_3Bi_5$ TDO measurement in the low and high frequency range respectively. **(f,h)** $1/\cos(\theta)$ fitting for the angular dependence of the Fourier transform frequency for the observed fundamental frequency peaks for the $CsTi_3Bi_5$ TDO measurement in the low and high frequency range respectively.

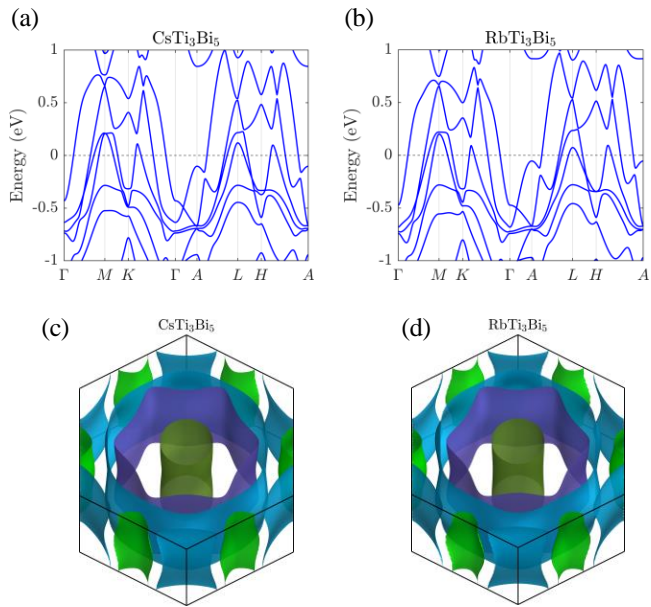

Figure 7. (a) and (b) are the band structures of $CsTi_3Bi_5$ and $RbTi_3Bi_5$, respectively, with spin-orbital coupling. (c) and (d) are the respective Fermi surfaces. Both the band structure and Fermi surface of $RbTi_3Bi_5$ are very similar to those of $CsTi_3Bi_5$. Notice that only the green Fermi surfaces around the $M - L$ lines are hole pockets while others are electron pockets.

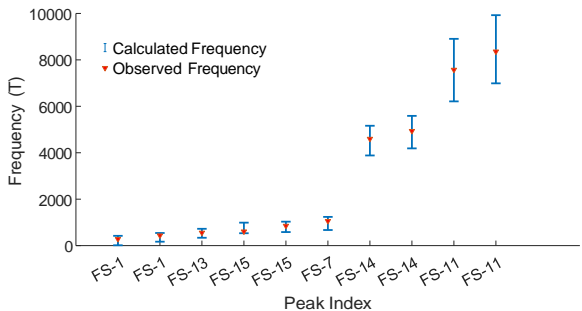

Figure 8. Calculated and observed Fourier transform peak frequencies for $CsTi_3Bi_5$ are shown with the Fourier peak index from the DFT calculations on the x-axis. The frequency dispersion per peak was calculated by tuning the Fermi level from -100 meV to 100 meV, with the error bars being the ±100 meV values of the Fermi level